\documentclass[preprint]{aastex}
\usepackage{natbib}
\shorttitle{}
\shortauthors{Nakashima et al.}

\begin{document}

\title{Discovery of the recombining plasma in the south of the Galactic center; a relic of the past Galactic center activity?}

\author{S. Nakashima, M. Nobukawa, H. Uchida, T. Tanaka, T. G. Tsuru and K. Koyama}
\affil{Department of Physics, Graduate School of Science, Kyoto University, Kitashirakawa Oiwake-cho, Sakyo-ku, Kyoto 606-8502, Japan}
\email{shinya@cr.scphys.kyoto-u.ac.jp}
\author{H. Murakami}
\affil{Department of Information Science, Faculty of Liberal Arts, Tohoku Gakuin University 2-1-1 Tenjinzawa, Izumi-ku, Sendai, Miyagi 981-3193}
\author{H. Uchiyama}
\affil{Science Education, Faculty of Education, Shizuoka University, 836 Ohya, Suruga-ku, Shizuoka,  422-8529, Japan}

\begin{abstract}
We report $Suzaku$ results for soft X-ray emission to the south of the Galactic center (GC).
The emission (hereafter  ``GC South'') has an angular size of $\sim 42\arcmin \times 16\arcmin$ centered at $(l,\ b)\sim(0.\arcdeg0,\ -1.\arcdeg4)$, 
and is located in the largely extended Galactic ridge X-ray emission (GRXE). 
The X-ray spectrum of GC South exhibits emission lines from highly ionized atoms.
Although the X-ray spectrum of the GRXE can be well fitted with a plasma in collisional ionization equilibrium (CIE), 
that of GC South cannot be fitted with a plasma in CIE, leaving hump-like residuals at $\sim$2.5 and 3.5~keV, 
which are attributable to the radiative recombination continua of the K-shells of Si and S, respectively.
In fact, GC South spectrum is well fitted with a recombination-dominant plasma model; 
the electron temperature is $0.46$~keV while atoms are highly ionized ($kT = 1.6$~keV) in the initial epoch, 
and the plasma is now in a recombining phase at a relaxation scale (plasma density $\times$ elapsed time) of $5.3\times10^{11}$~s~cm$^{-3}$. 
The absorption column density of GC South is consistent with that toward the GC region. 
Thus GC South is likely to be located in the GC region ($\sim$8~kpc distance). 
The size of the plasma, the mean density, and the thermal energy are estimated to be $\sim97~{\rm pc} \times 37~{\rm pc}$, 0.16~cm$^{-3}$, and $1.6 \times 10^{51}$~erg, respectively.
We discuss possible origins of the recombination-dominant plasma as a relic of past activity in the GC region. 
\end{abstract}


\section{Introduction}
The Galactic center (GC) region is unique  
because of its extreme environments, such as high-temperature gas, a high density of stars and clouds, and strong magnetic fields.
One of its characteristic features is the strong \ion{Fe}{25}~K$\alpha$ and \ion{Fe}{26}~K$\alpha$ emission, 
which indicates the presence of an extremely high temperature ($\sim$7~keV) plasma.
The plasma emission is unresolved and smoothly distributed over the GC region. 
Although the origin of the plasma, whether it is truly diffuse plasma or a superposition of faint point sources, is still under debate,  
past nuclear starburst activity is one of the possible origins based on what has found in other starburst galaxies (e.g., \citealt{2011ApJ...742L..31M}). 

At the dynamical center of our Galaxy, the supermassive black hole Sagittarius (Sgr) A*, whose mass is estimated to be $\sim$$4\times10^{6}~M_{\sun}$ \citep{2008ApJ...689.1044G}.
However, unlike an extragalactic active nucleus, the X-ray luminosity of Sgr\,A* is only $\sim$$10^{33-35}$~erg~s$^{-1}$,
which is $10^{9-11}$ times lower than the Eddington luminosity \citep{2003ApJ...591..891B,2003A&A...407L..17P}.  
On the other hand, several observations suggest that Sgr\,A* was active in the past. 
For example, the strong and time-variable \ion{Fe}{1} K$\alpha$ emissions from dense molecular clouds are interpreted 
as reflection and fluorescence due to strong X-rays from past Sgr\,A* flares \citep{2009PASJ...61S.241I, Ponti:2010hc, 2011ApJ...739L..52N}.
\cite{Ryu:2013aa} demonstrated that Sgr\,A* was continuously active 100--500~yr ago and suddenly became quiescent in the last 100~yr.   

Recent GeV gamma-ray observations revealed a giant bubble structure (the so-called ``{\it Fermi} bubbles''), 
which extends up to $50\arcdeg$ above and below the GC with a width of $40\arcdeg$  \citep{2010ApJ...724.1044S}.
This outstanding bipolar emission could have be formed by a past jet from Sgr\,A* or nuclear starburst activity.
Such outflow-like emissions were also reported at other wavelengths: 
the Wilkinson Microwave Anisotropy Probe haze \citep{Finkbeiner:2004vn}, the north polar spur \citep{1997ApJ...485..125S}, and the GC radio lobe \citep{2010ApJ...708..474L}.

If such energetic outflows were generated by past Sgr\,A* activity, then another relic might remain in the vicinity.
The {\it ROSAT} All Sky Survey has detected a smaller extended (plume-like) X-ray emission 1\arcdeg.5 south of Sgr\,A* (Wang 2002).
Although the physical properties of the plume-like emission have not been investigated yet, it may be related to past activity of the GC region. 
We thus performed observations of this region with {\it Suzaku} \citep{2007PASJ...59S...1M}.

In this paper, we adopt 8~kpc as the distance to the GC \citep{2008ApJ...689.1044G}.
At this distance, $1\arcmin$ corresponds to 2.3~pc. Errors are quoted at the 90\% confidence level.
 
\section{Observations and Data Reductions}
We observed the south of the GC using the X-ray imaging spectrometer (XIS; \citealt{Koyama2007}) on board {\it Suzaku} from 2009 March to 2011 March. 
Details of the observations are listed in Table \ref{tab:obslist}.
The XIS consists of three front-illuminated (FI) CCD cameras (XIS\,0, 2, and 3) and one back-illuminated (BI) CCD camera (XIS\,1).
The FI and BI CCDs have superior responses in the hard and soft bands, respectively.
The entire region of XIS\,2 and a part of XIS\,0 have not been functional because of anomalies, and were not used in the analysis.
Combined with the X-ray telescope (XRT; \citealt{2007PASJ...59S...9S}), the XIS covers a $17\arcmin.8 \times 17\arcmin.8$ field of view with a $2\arcmin$ angular resolution in the half-power diameter.
The effective areas of XIS with XRT are 330~cm$^{2}$ (FI) and 370~cm$^{2}$ (BI) at 1.5~keV.
The spaced-row charge injection (SCI) technique was applied to the XIS \citep{Bautz:2007tg} in order to repair the degradation of the charge transfer efficiency (CTE) due to radiation damage in orbit.
In the SCI mode, the energy resolutions (FWHMs) were 170~eV (FI) and 230~eV (BI) at 5.9~keV in 2010 after CTE calibration \citep{2009PASJ...61S...9U}.
All the observations were conducted in the normal clocking mode without window/burst options.

We used the software package HEASoft version 6.12.
After reprocessing with the calibration database released in 2011 October 10,  we screened the data to remove events during the South Atlantic Anomaly passages, with night-Earth elevation angles below 5\arcdeg and day-Earth elevation angles below 20\arcdeg.

\begin{deluxetable}{llcccc}
\tabletypesize{\scriptsize}
\tablecaption{Observation list. \label{tab:obslist}}
\tablewidth{0pt}
\tablehead{
\colhead{Seaquance \#} &  \multicolumn{2}{c}{Aim Point} & \colhead{Obs. Date} & \colhead{Exposure} & \colhead{ID in Fig. 1}  \\
& $\alpha_{2000.0}$ & $\delta_{2000.0}$ & \colhead{(yyyy-mm-dd)} & \colhead{(ks)}}
\startdata
503081010 &  268\arcdeg.05 & $-$29\arcdeg.76 & 2009-03-09 & 59 & \#1\\ 
504050010 &  267\arcdeg.86 & $-$29\arcdeg.58 & 2010-03-06 & 100 & \#2\\
504089010 &  267\arcdeg.56 & $-$29\arcdeg.60 & 2009-10-09 & 55 & \#3\\
505078010 &  267\arcdeg.95 & $-$29\arcdeg.80 & 2011-03-04 & 51 & \#4 \\
\cutinhead{Observations of the reference regions}
504002010  & 266\arcdeg.06 & $-$30\arcdeg.55 & 2010-02-27 & 53 & \#5 \\
504003010  & 266\arcdeg.39 & $-$30\arcdeg.62 & 2010-02-25 & 51 & \#6 \\ 
504090010  & 266\arcdeg.68 & $-$30\arcdeg.83 & 2009-10-13 & 41 & \#7 \\ 
504091010  & 267\arcdeg.09 & $-$31\arcdeg.05 & 2009-09-14 & 51 & \#8
\enddata
\tablecomments{Only the observations used in the spectral analysis are listed.}
\end{deluxetable}

\section{Analysis and Results}
The non--X-ray background (NXB) data were compiled using  {\tt xisnxbgen}, and  were subtracted from raw images and spectra.  
We used the SPEX software version 2.04.00 \citep{1996uxsa.conf..411K} in our spectral analysis. 
Redistribution matrix files and ancillary response files  were generated by {\tt xisrmfgen} and {\tt xissimarfgen}, respectively.
Spectra of XIS\,0 and XIS\,3 (FIs) were co-added since they have nearly the same responses.
The energy band affected by the contamination ($<$0.7~keV) and that around the neutral Si K-shell edge (1.7--1.8 keV) were ignored because of calibration uncertainties. 

\subsection{X-ray images in the south of the Galactic center}
\label{sec:image}
\begin{figure*}
\epsscale{1.0}
\plotone{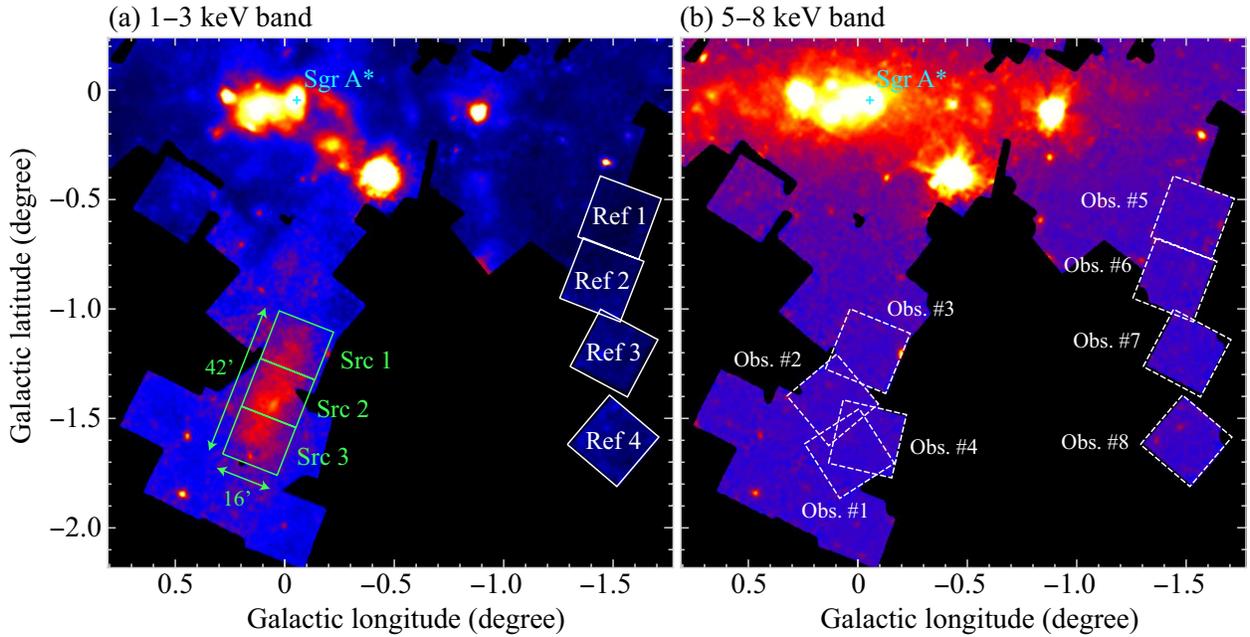}
\caption{
X-ray mosaic images of the region to the south of the GC.  
(a) The 1--3 keV band with the source and reference regions shown by the green rectangles and white squares, respectively.
(b) The 5--8 keV band with the observation fields listed in Table\ref{tab:obslist}.
The cyan cross indicates the position of  Sgr\,A*.
The vignetting effect was corrected after subtraction of the NXB. 
The data of three CCDs (XIS\,0, 1, and 3) were co-added.  
The data points were binned with $12\times12$ pixels and smoothed with a Gaussian kernel of $\sigma= 48\arcsec$. 
 \label{fig:GCimage}}
\end{figure*}

Figure \ref{fig:GCimage} shows the mosaic XIS images of the region to the south of the GC. 
The vignetting effect of the XRT was already corrected. 
The $Suzaku$ GC survey data were used to produce the images, although they are not included in Table\ref{tab:obslist} (see \citealt{Uchiyama:2013dq} for details).
In the 1.0--3.0 keV band, we found a significant excess over the Galactic background (BG) emission, which is largely extended as can be seen in the 5.0--8.0 keV band image. 
The excess has a center-filled morphology elongated from the north to the south and a size of $\sim 42\arcmin \times 16\arcmin$ centered at $(l,\ b)\sim(0.\arcdeg0,\ -1.\arcdeg4)$.
The shape and position are the same as those of the plume-like emission detected by {\it ROSAT} \citep{2002astro.ph..2317W}.
Hereafter, we refer to this excess as ``GC South''.

\subsection{background spectrum}
\label{sec:back}

\begin{figure*}
\epsscale{1.0}
\plotone{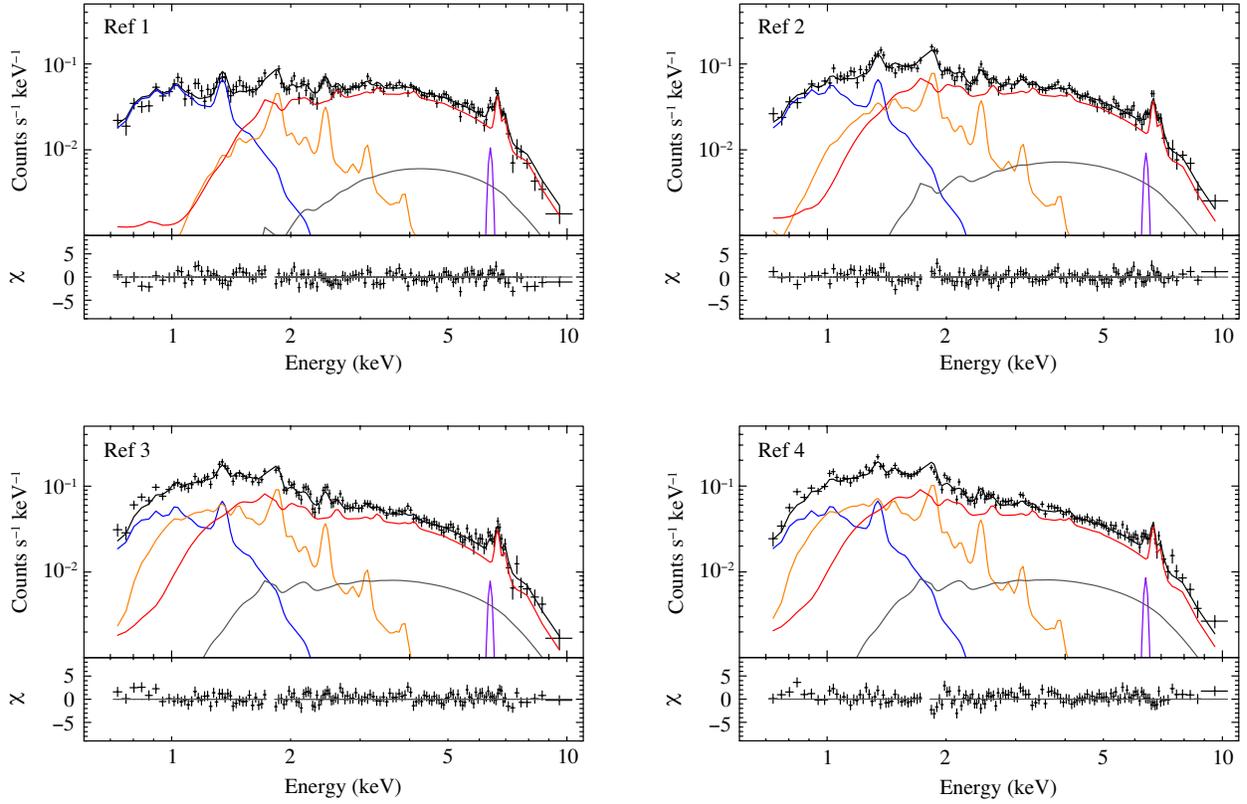}
\caption
{
Spectra and best-fit models of each reference region.
Only the co-added FI spectra and models are displayed for visibility, although the FI and BI spectra were simultaneously fitted. 
The red, orange, purple, blue, and gray lines represent HP, LP, Gaussian, FE, and CXB, respectively (see the text).
}\label{fig:bgdspec}
\end{figure*}

X-ray sources in the GC region are affected by the strong Galactic BG, whose X-ray flux depends on position with various components. 
Thus, careful evaluation of the BG emission is essential.
\citet{2011PASJ...63S.903U,Uchiyama:2013dq} extensively studied this emission with {\it Suzaku}, 
and determined the spatial distribution along the Galactic latitude and longitude.  
On the based of the spatial distribution, they divided the BG emission into two components: 
the Galactic center X-ray emission (GCXE) and the Galactic ridge X-ray emission (GRXE).  
\cite{Uchiyama:2013dq} reported that the GCXE is elliptical with a size of $\sim0.6\arcdeg \times 0.3\arcdeg$ while the GRXE is at least one order of magnitude larger scale.
Therefore, GC South is located in the region where contributions from the GRXE are dominant.

We extracted X-ray spectra from the four reference regions shown in Figure \ref{fig:GCimage} (Ref\,1--4).
They show many emission lines of various elements that are typical of the GRXE.  
We therefore constructed a BG model based on the GRXE model reported by \cite{Uchiyama:2013dq}. 
The model includes a high temperature plasma component (HP), a low temperature plasma component (LP), and a foreground emission component (FE),
all of which are modeled by collisional ionization equilibrium (CIE) plasmas. 
Although the FE consists of 0.59~keV and 0.09~keV plasmas in the study of \cite{Uchiyama:2013dq}, 
we applied a single $\sim$0.59~keV plasma because the  contribution of the 0.09~keV plasma is negligible in the energy range used in our analysis (0.7--10 keV).  
Free parameters include the temperatures ($kT_{\rm HP}$,  $kT_{\rm LP}$,  $kT_{\rm FE}$ ) and the emission measures ($EM_{\rm HP}$, $EM_{\rm LP}$, $EM_{\rm FE}$).
The metal abundances relative to solar values \citep{1989GeCoA..53..197A} were common between the HP and LP.
Those of Mg, Si, S, Ar, Ca, and Fe were free parameters while the Ne and Ni abundances were linked to Mg and Fe, respectively.
The other elements were fixed to the solar values.
For the FE, the metal abundances of Ne and Mg were free parameters while those of the other elemetns were linked and allowed to vary. 
In order to reproduce the \ion{Fe}{1} K$\alpha$ emission in the spectra, we also added a Gaussian component with a fixed line centroid of 6.4 keV.
The continuum emission associated with the \ion{Fe}{1} K$\alpha$ emission was included in the HP.
In addition, we need to include the cosmic X-ray background (CXB) even though its contribution is small. 
The CXB was represented by a single power-law function with a photon index and surface brightness (2--10~keV) of 1.41 and $6.38 \times 10^{-8}$~erg~cm$^{-2}$~s$^{-1}$~sr$^{-1}$, respectively \citep{2002PASJ...54..327K}.
The explicit form of our BG model is therefore
\begin{equation}
{A_ {\rm GC}} \times ({\rm HP}+{\rm LP}+ {\rm Gaussian}) + {A_{\rm FE}} \times {\rm FE} + (A_{\rm GC})^{2} \times {\rm CXB},
\label{eq:model1}
\end{equation}
where ${\rm A_{GC}}$ (column density of $N_{\rm H\,GC}$) and ${\rm A_{FE}}$ (column density of $N_{\rm H\,FE}$) represent the interstellar absorptions toward the GC and for the FE component, respectively.
The absorption for the CXB is twice that of $N_{\rm H\,GC}$, because the CXB is of extragalactic origin behind the GC region.

We simultaneously fitted the reference-region spectra with this model.
Since the flux in the term (HP + LP + Gaussian) depends on the position as shown in \cite{Uchiyama:2013dq},
we multiplied by a constant factor as
\begin{equation}
{A_{\rm GC}} \times ({\rm HP}+{\rm LP}+ {\rm Gaussian}) \times factor,
\label{eq:model2}
\end{equation}
where $factor$ for Ref\,1 was fixed to 1, and those for the other regions were free parameters.
The $N_{\rm H\,GC}$ values were also independently allowed to vary while the other parameters were common among the regions.
As a result of our fitting, the spectra were well reconstructed with  $ \chi^{2}/{\rm d.o.f} = 1415/1107$. 
The best-fit models and parameters are shown in Figure \ref{fig:bgdspec} and Table \ref{tab:bgdfit}, respectively. 
We note that the HP, LP and FE parameters are slightly different from those of \cite{Uchiyama:2013dq}.  
This is due to the difference of the models and selected regions; the GRXE spectrum of \cite{Uchiyama:2013dq} is a composite from the on-plane regions of $|l| >2\arcdeg.0$ and $|b| <0\arcdeg.5$, while our BG regions are far from the plane of $l =-1\arcdeg.4$ and $-1\arcdeg.6  < b < -0\arcdeg.6 $. 
One might claim that the abundances of some elements are significantly smaller than the solar values. 
We approximated the BG spectra using simplified plasma models based on the GRXE model near the GC \citep{Uchiyama:2013dq}. 
In the Galactic bulge regions (our observation fields) however, the possible contribution of unresolved faint point sources (e.g. \citealt{2009Natur.458.1142R}), 
CXB fluctuation \citep{2002PASJ...54..327K}, or unknown non thermal emissions may not be ignored. 
These sources increase the continuum flux 
in the BG spectrum, and hence reduce the apparent abundances.
In order to verify this possibility, we fitted the spectra by fixing the abundances of all the elements in the relevant plasma models (LP, HP, and FE) to the solar values, 
and adding two power-law components; one is convolved with $A_{\rm{GC}}$, and the other is convolved with $A_{\rm{FE}}$.
Then we obtained a reasonable fit ($\chi^{2}/{\rm d.o.f} = 1654/1116$) with no significant change for all the parameters in Table \ref{tab:bgdfit}.

\begin{figure}
\epsscale{1.0}
\plotone{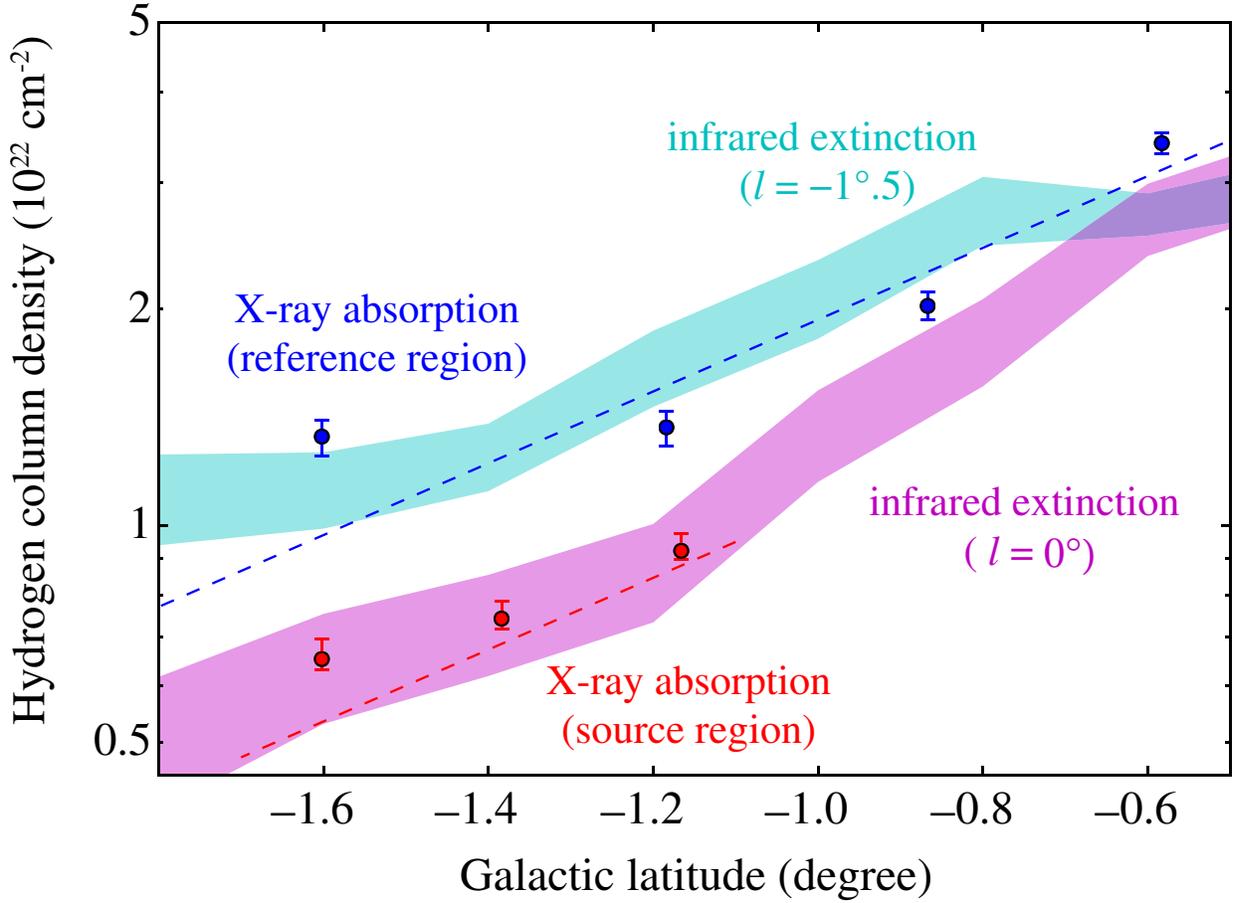}
\caption{
Profile of the absorption column density. 
The blue circles indicate $N_{\rm H\,GC}$ obtained from the best-fit  of the reference regions.
The broken blue  line is the exponential-folding model of $N_{\rm H\,GC}$ (see the text).
The blue and red shaded areas show the near-infrared extinctions ($A_{K_{\rm s}}$) including the errors along $l=-1\arcdeg.5$ and $l=0\arcdeg$, respectively \citep{2012A&A...543A..13G}. 
The  near-infrared extinctions were converted to $N_{\rm H}$ with the scaling factor of $N_{\rm H}/A_{K_{\rm s}} = 1.8\times 10^{22}$~cm$^{-2}$~mag$^{-1}$. 
The red circles indicate the fitting results of the spatially resolved source spectra.
\label{fig:absorption}}
\end{figure}

The best-fit $factor$ values decrease with increasing the Galactic latitude ($b$) with an exponential-folding scale of  $3\arcdeg.3\pm0\arcdeg.8$, which is roughly consistent with those of \cite{Uchiyama:2013dq}.
The $N_{\rm H\,GC}$ values also decrease with $b$, as phenomenologically given by
\begin{equation}
N_{\rm H\,GC}  
=  6.1\pm0.7\times10^{22}\exp{\left(-\frac{|b|}{0\arcdeg.87\pm0\arcdeg.10}\right)}~{\rm cm}^{-2}.
\label{eq:nh}
\end{equation}
This model and the best-fit $N_{\rm H\,GC}$ are plotted in Figure \ref{fig:absorption} by the blue broken line and circles, respectively.
We compared these data with the near-infrared extinction ($A_{K_{\rm s}}$) of the stars located in the GC (\citealt{2012A&A...543A..13G}).
The mean $A_{K_{\rm s}}$ and best-fit $N_{\rm H\,GC}$ in the region of $ -1\arcdeg.6 < b < -1\arcdeg.2$  are 0.76~mag  and $1.35\times10^{22}$~cm$^{-2}$, respectively.  
Thus, a conversion factor $N_{\rm H}/A_{K_{\rm s}}$ was estimated to be  $1.8 \times 10^{22}$~cm$^{-2}$~mag$^{-1}$,  
which is within the range reported in the literature (e.g., \citealt{2009ApJ...700.1702V,2009MNRAS.400.2050G}).
With this conversion factor, the $A_{K_{\rm s}}$ profile is superimposed on Figure \ref{fig:absorption} as the blue shaded area. 
We confirmed that the $N_{\rm H\,GC}$ model well follows the over-all trend of the $A_{K_{\rm s}}$ profile well.
  
Figure \ref{fig:absorption} also shows $A_{K_{\rm s}}$ in the source region ($l=0\arcdeg.0$) with the red shaded area  \citep{2012A&A...543A..13G}.
We found that the $A_{K_{\rm s}}$ in the source region is smaller than that in the reference region by a factor of $\sim$0.55 in the $ -1\arcdeg.6 < b < -1\arcdeg.2$ range.
Assuming  the same $b$-dependence as in the reference regions, we estimated $N_{\rm H\,GC}$  in the source region by normalizing Equation (\ref{eq:nh}) by a factor of 0.55.
This estimated model is shown by the broken red line in Figure \ref{fig:absorption}.

\begin{deluxetable}{lllll}
\tabletypesize{\scriptsize}
\tablecaption{Fitting result of the spectra in the reference regions.\label{tab:bgdfit}}
\tablewidth{0pt}
\tablehead{\colhead{Parameter (Unit)}  & \colhead{Ref\,1} & \colhead{Ref\,2} & \colhead{Ref\,3} & \colhead{Ref\,4}}
\startdata
\multicolumn{5}{l}{--- HP + LP + Gaussian ---} \\
$N_{\rm{H\,GC}}$ ($10^{22}$ cm$^{-2}$) & $3.40\pm0.11$ & $2.02\pm0.09$ & $1.37\pm0.08$ & $1.33\pm0.08$ \\
$factor$ & 1 (fixed) & $0.85\pm0.03$ & $0.71\pm0.02$ & $0.77\pm0.02$ \\
\\
$kT_{\rm HP}$ (keV) &  \multicolumn{4}{c}{$7.6^{+0.4}_{-0.3}$} \\
$kT_{\rm LP}$ (keV) &  \multicolumn{4}{c}{$0.69\pm0.03$} \\
$EM_{\rm HP}\tablenotemark{a}$ ($10^{58}$ cm$^{-3}$) &  \multicolumn{4}{c}{$0.38\pm0.01$}\\
$EM_{\rm LP}\tablenotemark{a}$ ($10^{58}$ cm$^{-3}$) &  \multicolumn{4}{c}{$0.45^{+0.09}_{-0.08}$}\\
Mg = Ne (solar) &  \multicolumn{4}{c}{$0.43^{+0.14}_{-0.12}$} \\
Si (solar) &  \multicolumn{4}{c}{$0.94^{+0.17}_{-0.14}$} \\
S (solar) &  \multicolumn{4}{c}{$1.6\pm0.3$} \\
Ar (solar) &  \multicolumn{4}{c}{$3.5\pm0.8$} \\
Ca (solar) &  \multicolumn{4}{c}{$2.8\pm1.0$} \\
Fe = Ni (solar) &  \multicolumn{4}{c}{$0.53\pm0.04$} \\
Other Elements (solar) &  \multicolumn{4}{c}{1 (fixed)} \\
\ion{Fe}{1} K$\alpha$ flux ($10^{-6}$~ph~cm$^{-2}$~s$^{-1}$) &  \multicolumn{4}{c}{$6.4\pm1.2$} \\
\\
\multicolumn{5}{l}{--- FE ---} \\
$N_{\rm{H\,FE}}$ ($10^{22}$ cm$^{-2}$) &  \multicolumn{4}{c}{$0.42\pm0.01$} \\
$kT_{\rm FE}$ (keV) &  \multicolumn{4}{c}{$0.33\pm0.04$} \\
$EM_{\rm FE}\tablenotemark{a}$ ($10^{58}$ cm$^{-3}$) &  \multicolumn{4}{c}{$0.86^{+0.66}_{-0.33}$}\\
Ne (solar) &  \multicolumn{4}{c}{$0.06\pm0.02$} \\
Mg (solar) &  \multicolumn{4}{c}{$0.19^{+0.05}_{-0.04}$} \\
Other Elements (solar) &  \multicolumn{4}{c}{$0.022^{+0.014}_{-0.009}$}
\enddata
\tablenotetext{a}{Emission measure assuming the distance of 8~kpc. $EM = \int n_{\rm{e}}n_{\rm{p}}dV$, where $n_{\rm{e}}$, $n_{\rm{p}}$, and V , are the electron density, the proton density, and the emitting volume, respectively.}
\end{deluxetable}

\subsection{Spectral analysis of the entire region}
\label{sec:merge}
\begin{figure}
\epsscale{1.0}
\plotone{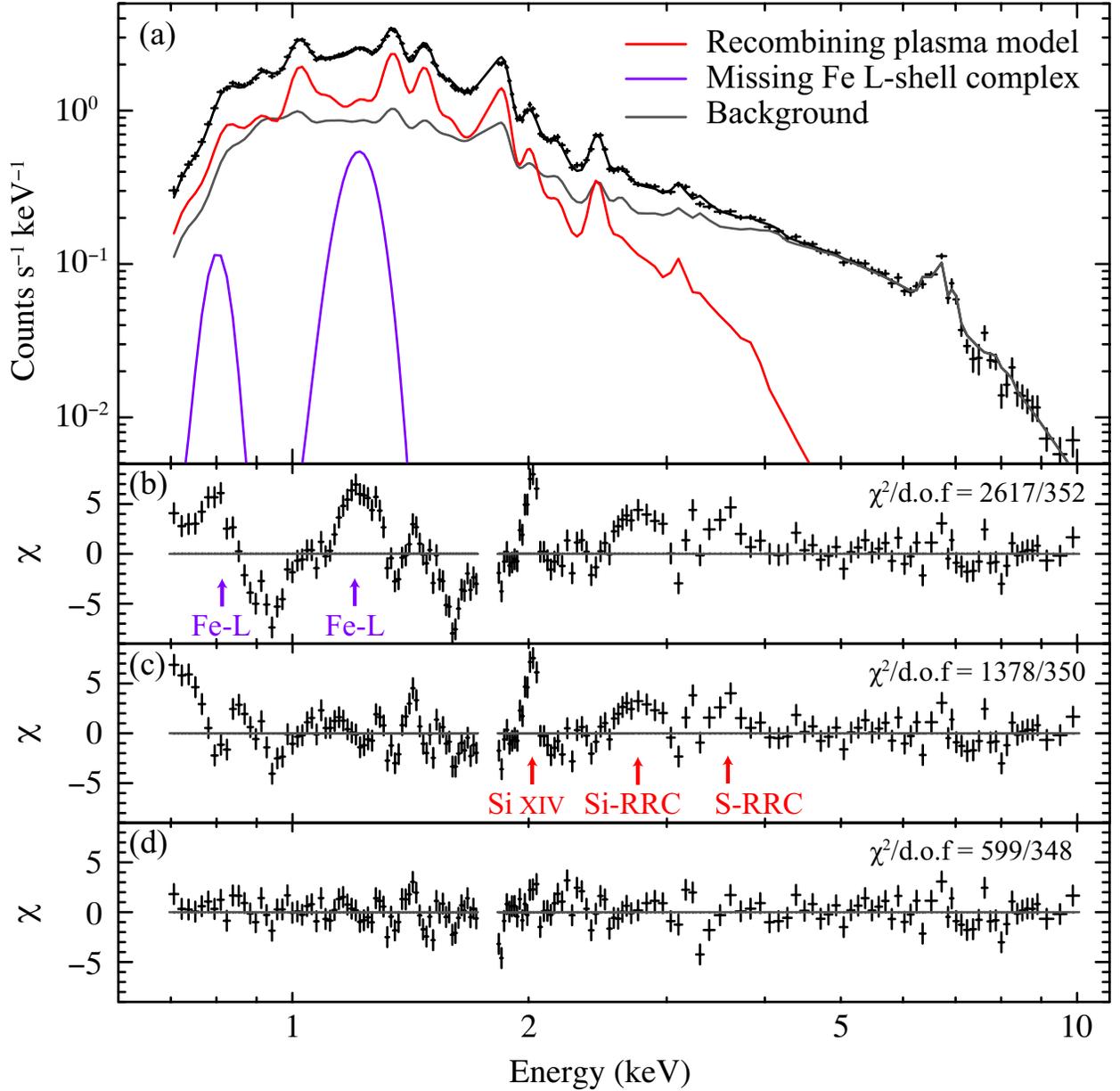}
\caption{
(a) Spectrum extracted from the entire source region. Only the co-added FI spectra and models are displayed for visibility, although the FI and BI spectra were simultaneously fitted. The solid red, purple, and gray lines indicate the best-fit RP model, missing Fe L-shell lines, and the background model, respectively.
The $\chi^{2}$ values fitted with each of the models are also shown: (b) the CIE plasma model, (c) the CIE plasma and additional Fe L-shell transient model, and (d) the RP and additional Fe L-shell transient model (best fit).
\label{fig:MergedSpec}}
\end{figure}

We first investigated GC South spectrum extracted from the entire region ($42\arcmin \times 16\arcmin$ rectangular region) 
in order to increase the photon statistics and to reveal the overall physical properties. 
The spectrum is plotted in Figure \ref{fig:MergedSpec}(a).  
The hard X-ray band above 5~keV is dominated by the BG emission as shown in Figure \ref{fig:GCimage}(b).  
In order to subtract these BG components, we applied the BG model with the best-fit values obtained from the reference regions (Section \ref{sec:back}).   
Only the surface brightness and $N_{\rm H\,GC}$ for the (HP + LP + Gaussian) component should be different from those in the reference regions.
Therefore, $factor$ was treated as a free parameter, and the $N_{\rm H\,GC}$ value was fixed to $0.69\times10^{22}$ cm$^{-2}$, 
which is the mean value of the $N_{\rm  H\,GC}$ profile in the source regions (the broken red line in Figure \ref{fig:absorption}).
The parameters for the FE were the same as those in the reference regions.

The BG model with these parameters reproduced the 5--10 keV band spectrum very well,
but it left a large data excess in the 0.7--5.0 keV band.  
This excess shows emission lines from highly ionized atoms, and hence is of thermal plasma origin.  
We then attempted a fitting with a model of combining CIE plasma and the BG  model. 
The free parameters in the CIE plasma were $N_{\rm H}$, $kT$, and the abundances of the major  
elements (Ne, S, Si, Ar, and Fe). 
The abundances of  O, Ca, and Ni were linked to those of Ne, Ar, and Fe, respectively.  
Those of the other elements were fixed to the solar values. 

Spectral fitting with the composite model gave a best-fit CIE temperature of $kT=0.62$~keV,  
but failed to reconstruct the data with $\chi^{2}/{\rm d.o.f}=2617/352$ (Figure \ref{fig:MergedSpec}(b)).
The most prominent residuals in Figure \ref{fig:MergedSpec}(b) are the line and bump at 
0.8~keV and 1.2--1.3~keV, respectively.  These are known to originate from incomplete atomic data 
for the Fe L-shell complex in the current plasma model.
\citet{2007ApJ...670.1504G} claimed that the line intensity ratio of 
3s$\rightarrow$2p  ($\sim$0.7~keV) over 3d$\rightarrow$2p ($\sim$0.8~keV) for \ion{Fe}{18} has an uncertainty.
\citet{2000ApJ...530..387B} and \citet{2001A&A...365L.329A} reported that the SPEX 
plasma code underestimates the flux around 9.6--10.6~{\AA} (1.17--1.29 keV), 
since some Fe L-shell transitions ($n=6,7,8\rightarrow2$ for \ion{Fe}{17}, 
$n=6,7\rightarrow2$ for \ion{Fe}{18}, and $n=6\rightarrow2$ for \ion{Fe}{19}) are missing.
We thus phenomenologically added Gaussians at 0.8~keV and 1.2~keV.
The FWHM of the 1.2~keV line was fixed to 120~eV to approximate multiple lines.
Then, the fit was improved with $\chi ^{2}/{\rm d.o.f}=1378/350$, but large residuals still remained (Figure \ref{fig:MergedSpec}(c)). 
We conclude that the CIE model for GC South plasma is invalid.

In Figure \ref{fig:MergedSpec}(c),  the data excess  at 2.0~keV  corresponds to  \ion{Si}{14} K$\alpha$.  
This indicates that the CIE model ($kT=0.62$~keV) underestimates the ionization state of Si. 
In addition, we can clearly observe hump-like features around 2.5--3.0~keV and 3.2--3.8~keV. 
These features are most likely due to the radiative recombination continua (RRCs) of Si and S, which are observed in some Galactic supernova remnants (SNRs; see \citealt{2012PASJ...64...81S,Uchida:2012aa} and references therein).
Although the residual of \ion{Si}{14} K$\alpha$ can be solved by a multi-temperature CIE model, the RRCs cannot be reproduced. 

The RRC residuals strongly indicate that the plasma is in the recombining phase.
We thus applied a non-equilibrium ionization model ({\tt NEIJ} in SPEX), 
that successfully reproduces the RRC features observed in the SNRs W44 and G346.6$-$0.2 \citep{Uchida:2012aa,Yamauchi:2013cr}.
This model traces the evolution of the ionization state 
when the initial ionization temperature ($kT_{\rm init}$) and the electron 
temperature ($kT_{\rm e}$, assumed to be constant) are given;
the following evolution of the ionization state is traced by the relaxation timescale 
of $n_{\rm{e}}t$, where $n_{\rm{e}}$ and $t$ are the number density of electrons and the elapsed time, respectively.  
If the best-fit $kT_{\rm init}$ is $\sim$0, then this model describes an ionizing plasma ({\tt NEI} in XSPEC), 
while if $kT_{\rm init} > kT_{\rm e}$, it describes a recombining plasma (RP).

The free parameters for the {\tt NEIJ} model were $kT_{\rm init}$, $kT_{\rm e}$, $n_{\rm{e}}t$,  
and the abundances of the major elements (the same as in the CIE fitting).
The {\tt NEIJ} model gave a good fit with $\chi ^{2}/{\rm d.o.f}=599/348$.
The best-fit $kT_{\rm init}$, $kT_{\rm e}$, and $n_{\rm{e}}t$  are 1.6~keV, 0.46~keV, and  $5.3\times10^{11}$~s~cm$^{-3}$ , respectively (see Table \ref{tab:srcfit}). 
This best-fit model well reproduced the RRC features around 2.5--3.0~keV and 3.2--3.8~keV, as shown in Figure \ref{fig:MergedSpec}(d).

For comparison, we list the average charge state of each element in the best-fit {\tt NEIJ}  
model and that of the CIE at the same electron temperature (Table \ref{tab:charge}). 
We confirmed that GC South is really an RP.

\subsection{Analysis of the spatially resolved spectra}
\label{sec:resolve}
We found that the recombining process dominates in the plasma based on the entire source spectrum.
In order to study the spatial variation of the absorption column density and the ionization state,
we divided the source region into three (Src\,1, 2, and 3), as shown in Figure \ref{fig:GCimage}, and simultaneously fitted their spectra with the {\tt NEIJ} (RP) model.
For the BG model, the parameters of $factor$  were treated as free parameters in the same way as the entire spectrum fitting.
We used the $N_{\rm H\,GC}$ values given in Figure \ref{fig:absorption} as the  dashed red line. 

\begin{figure}
\epsscale{0.5}
\plotone{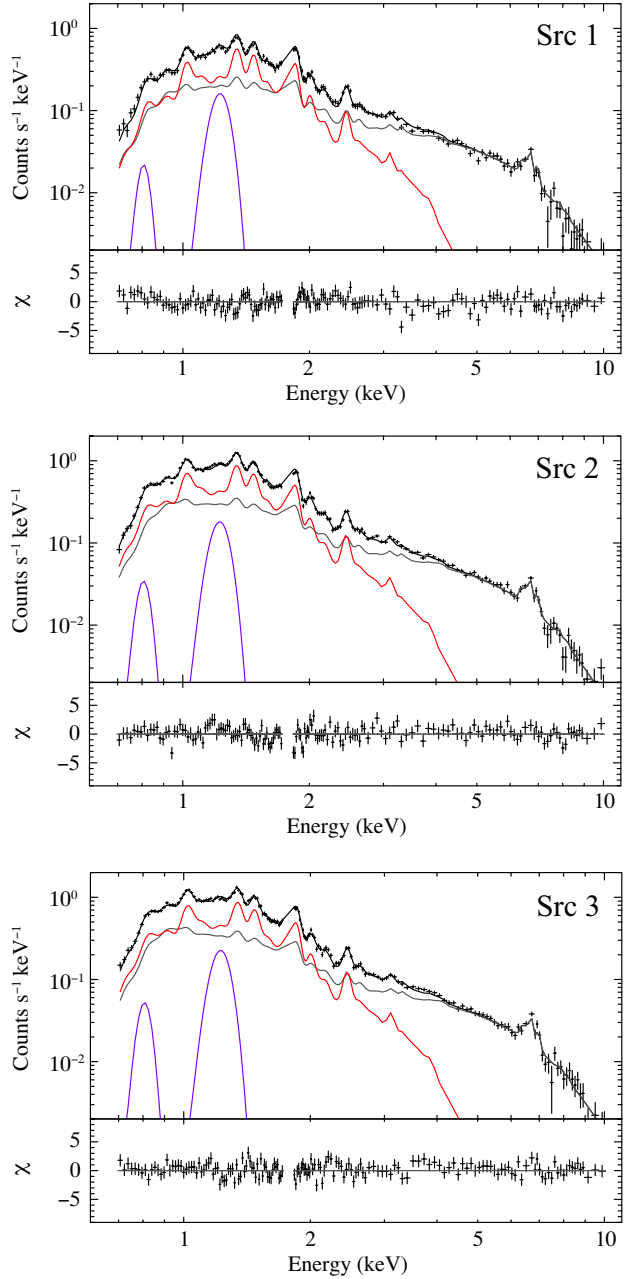}
\caption{Spatially resolved spectrum of the source. Only the co-added FI spectra and models are displayed for visibility, though the FI and BI spectra were simultaneously fitted. 
The red, purple, and gray lines represent the RP model, the missing Fe L-shell complex, and the background model, respectively.
\label{fig:divspec}}
\end{figure}

For the {\tt NEIJ} (RP) model, $kT_{\rm init}$ was fixed at 1.6~keV, which is the best-fit value for the entire spectrum, while $N_{\rm H}$,
$kT_{\rm{e}}$ and $n_{\rm{e}}t$ were allowed to vary independently and the other parameters were tied between the three regions. 
Then, the spectra were accurately reproduced with $\chi ^{2}/{\rm d.o.f}=1584/1186$, as shown in Figure \ref{fig:divspec}.
Using the best-fit results, we calculated the surface brightness in the 5--8 keV band for the three source regions and the four reference regions (Section \ref{sec:back}).  
The results, both the absolute fluxes and the ($l$, $b$) distributions, agree with those in Table 2 of \cite{Uchiyama:2013dq}.
This supports  the reliability of our method of BG estimation and subtraction.
The best-fit parameters for the source model ({\tt NEIJ}) are shown in Table \ref{tab:srcfit}.
The  absorption column densities ($N_{\rm H}$)  for the Src\,1, 2, and 3 are plotted 
in Figure \ref{fig:absorption} with red circles.
These are reasonably consistent with the estimated $N_{\rm H\,GC}$ profile (and that of the near-infrared extinction) at the same latitudes.
We found no spatial variation of the electron temperature or the relaxation
timescale. Therefore, the key physical parameters for the plasma are uniform within the whole area of GC South.

\begin{deluxetable}{lllll}
\tabletypesize{\scriptsize}
\tablecaption{Fitting results of the source spectrum with {\tt NEIJ}. \label{tab:srcfit}}
\tablewidth{0pt}
\tablehead{
\colhead{} & \colhead{} & \multicolumn{3}{c}{Spatially resolved fitting\tablenotemark{a}} \\
\cline{3-5} \\
\colhead{Parameters (Unit)} & \colhead{Entire region} & \colhead{Src\,1} & \colhead{Src\,2} & \colhead{Src\,3}
}
\startdata
$N_{\rm{H}}$ (10$^{22}$ cm$^{-2}$) & $0.70\pm0.03$ & $0.94\pm0.03$ & $0.77\pm0.03$ & $0.63^{+0.04}_{-0.02}$\\
$kT_{\rm{e}}$ (keV) & $0.46\pm0.02$ & $0.45^{+0.05}_{-0.03}$ & $0.47\pm0.02$ & $0.47\pm0.02$ \\ 
$kT_{\rm{init}}$ (keV) & $1.63^{+0.27}_{-0.18}$ & & $1.63$ (fixed) & \\
$nt$ (10$^{11}$ s cm$^{-3}$) & $5.3\pm0.5$ & $5.2^{+0.5}_{-0.4}$ & $5.7\pm0.3$ & $5.2\pm0.3$\\
$EM\tablenotemark{b}$ (10$^{58}$ cm$^{-3}$) & $9.5\pm1.0$ & $2.7^{+0.4}_{-0.5}$  & $3.6\pm0.4$ & $3.1^{+0.4}_{-0.3}$\\ 
Ne =O  (solar) & $0.39^{+0.05}_{-0.04}$ & & $0.44^{+0.07}_{-0.06}$ & \\
Mg (solar) & $0.81\pm0.07$ & & $0.86^{+0.09}_{-0.07}$ & \\
Si (solar) & $0.71\pm0.05$ & & $0.71^{+0.06}_{-0.05}$ & \\
S (solar) & $0.79^{+0.09}_{-0.08}$ & & $0.78^{+0.09}_{-0.08}$ & \\
Ar = Ca (solar) & $1.9^{+0.6}_{-0.5}$ & & $1.9\pm0.4$ & \\
Fe = Ni (solar) & $0.10\pm0.02$ & & $0.12\pm0.02$ & 
\enddata
\tablenotetext{a}{The parameters of $N_{\rm{H}}$,  $kT_{\rm{e}}$, $nt$, and $EM$ were independently allowed to vary among the three regions while the other parameters were common.
We fixed $kT_{\rm{init}}$ to 1.63 keV, which is the best-fit value in the entire-region fitting.
 }
\tablenotetext{b}{Emission measure assuming the distance of 8~kpc. $EM = \int n_{\rm{e}}n_{\rm{p}}dV$, where $n_{\rm{e}}$, $n_{\rm{p}}$, and V , are the electron density, the proton density, and the emitting volume, respectively.}
\end{deluxetable}

\begin{deluxetable}{lccccccc}
\tablecaption{Averaged charge state of each element \label{tab:charge}}
\tablewidth{0pt}
\tablehead{
\colhead{model} & \colhead{Ne} & \colhead{Mg} & \colhead{Si} & \colhead{S} & \colhead{Ar} & \colhead{Fe}  
}
\startdata
best-fit {\tt NEIJ} & 9.56 & 11.1 & 12.7 & 14.2 & 15.7 & 17.0\\
CIE ($kT = 0.46$~keV)& 9.30 & 10.3 & 12.0 & 13.9 & 15.6 &16.4 
\enddata
\end{deluxetable}

\section{Discusssion}

\subsection{Physical properties}
As shown in Figure \ref{fig:absorption}, absorption column densities for GC South (the red circles) are consistent with those toward the GC region (the broken red line). 
Thus, GC South is most likely to be located in the GC region (8~kpc distance). 
The size of the plasma is estimated to be 97~pc $\times$ 37~pc at this distance.
Assuming that the volume of the plasma is $97~{\rm pc} \times 37~{\rm pc} \times 
37~{\rm pc} = 1.3 \times 10^{5}~{\rm pc}^{3}$, and using a best-fit emission measure ($EM$) of 
$9.5\times10^{58}$ cm$^{-3}$, the electron density of the plasma is calculated to 
be 0.16~cm$^{-3}$.
Accordingly, the X-ray emitting mass and thermal energy are estimated to be 
$7.1\times10^{2}$~$M_{\sun}$ and $1.6\times10^{51}$~erg, respectively.
From the relaxation time scale ($nt = 5.3\times10^{11}$ s cm$^{-3}$), the age of 
the plasma is $\ga$$1.1\times10^{5}$~yr.

\subsection{Origin of the GC south plasma}
\label{sec:origin}
Typical thermal energies observed in Galactic X-ray SNRs are $\sim$$10^{49-50}$~erg.  
Thus, GC South is much more energetic than is expected to be produce by a single supernova (SN).  
In fact, we cannot find a non-thermal radio shell near GC South (e.g., \citealt{2000AJ....119..207L}).
Multiple SNe and stellar winds could provide sufficient energy ($>10^{51}$~erg) for the formation of the GC South like Galactic super bubbles.  
However, neither the OB association nor the \ion{H}{2} region has been detected in this region. 
Thus, a single or a multiple SN origin is unlikely.

One possible origin of GC South could be past starburst activity.
For example, the starburst galaxy M82 has a blob called the ``Cap''  at  the termination point of the superwind from the host galaxy (\citealt{2007PASJ...59S.269T}).  
The size and thermal energy of the Cap are $3.7 \times 0.9$~kpc and $\sim$$10^{55}$~erg, respectively.
Both are larger than those of GC South by two to four orders of magnitude.  
Thus, GC South might have been formed by small super-wind activity in our Galaxy.
Indeed,  \cite{Matsunaga:2011aa} and \cite{Yusef-Zadeh:2009aa} claimed that the star formation rate $10^{5-7}$ yr ago in the GC region was higher than the current rate 
by one order of magnitude ($\sim$0.1~$M_{\sun}$~yr$^{-1}$). 
Assuming a simple initial mass function \citep{1955ApJ...121..161S}, 100 SNe of massive stars ($>$8~$M_{\sun}$) are possible within $\sim$$10^{5}$ yr.
This number of SNRs is sufficient to provide the thermal energy of GC South.
If the plasma was formed in the GC and blown-out at the sound speed 
(510~km~s$^{-1}$ for 1.6 keV plasma), then $4\times10^{5}$~yr is required to reach
1\arcdeg.7 (230~pc) south of the GC.  This time scale is consistent with the plasma 
age, and hence is consistent with the starburst origin.

Another possibility is that GC South was created by Sgr A* activity $10^{5}$ yr ago. 
Such activity may be conceivable, because similar activity $\sim$300~yr and $\sim$$10^{6}$~yr ago has been suggested by \cite{1996PASJ...48..249K} and \cite{2010ApJ...724.1044S}, respectively.
The flare probably emitted a bipolar flow perpendicular to the Galactic plane, and generated a shock-heated hot plasma. 
If the speed of the bipolar flow is a few hundred km~s$^{-1}$, 
then the shocked plasma would be formed  230 pc away from Sgr A* with a sub-keV temperature, as is observed for the values of GC South. 

\subsection{Mechanism to form the recombining plasma}
We discovered that GC South is an RP, which is not predicted in the standard evolution of a shock-heated plasma;
a shock-heated low-density ($\sim$1~cm$^{-3}$) plasma is in an ionizing phase for a long period 
because the timescale of electron heating ($\sim$100~yr) is much shorter than that of collisional ionization ($\sim$$10^{4}$~yr).  
Therefore, to realize an RP, some specific events should occur: 
either a rapid decrease in the electron temperature or enhancement of ionization only.

In SNRs, two mechanisms for the decrease of the electron temperature have been proposed (e.g., \citealt{2012PASJ...64...81S,Uchida:2012aa}). 
Numerical simulations support these mechanisms (e.g., \citealt{Zhou:2011vn,2012PASJ...64...24S}). 
We attempt to apply these scenarios to GC South.
One is thermal conduction via interactions with ambient cold molecular clouds.  
However, no such molecular cloud has been found in this region.  
Moreover, no spatial gradient of the electron temperature is found in GC South as described in Section \ref{sec:resolve}. 
Therefore, this scenario is unlikely. 
The other scenario is an adiabatic expansion of the plasma.
The CIE plasma would have been formed in the dense circumstellar medium at the starburst site. 
If the plasma was blown out rapidly to a thin circumstellar space, then the electron temperature would be rapidly cooled down as a result of the adiabatic expansion.
In order to decrease the electron temperature from 1.6 keV to 0.46 keV, the plasma 
volume should increase by a factor of six, following the
$T\propto V^{1-\gamma}$ relation under the adiabatic condition ($\gamma = 5/3$ for a 
monatomic gas).
Assuming that the size of the plasma increases from 54~pc to 97~pc (a factor of $6^{1/3} = 1.8$) at a sound 
speed of 510~km~s$^{-1}$, the expansion time scale is calculated to be 
$\sim$$8\times10^{4}$~yr, which is smaller than  the observed recombination 
time scale of $1.1\times10^{5}$~yr.  Therefore, the adiabatic expansion is 
one possible scenario. 

Selective ionization also formes an RP.
One possibile scenario is photoionization.
In this scenario, a plasma is irradiated by a strong X-ray source and is over-ionized initialy.
After the termination of photoionization, the plasma enters a recombining phase.
The balance between photoionization and recombination is described  by a single parameter, $\xi = L/(nR^{2})$, 
where $L$, $n$, and $R$ are the luminosity of the source, the gas density, and the distance between the irradiating source and the plasma gas, respectively. 
In order to achieve nearly the same ionization state as the 1.6~keV plasma, 
$\xi$ should be $\sim$1000 (\citealt{2001ApJS..133..221K}).  
Then, the required luminosity is estimated as
\begin{equation}
L = 7.6 \times10^{43} \left(\frac{R}{230~{\rm pc}}\right)^{2} \left(\frac{n}{0.16~{\rm cm}^{-3}}\right)~{\rm erg~s}^{-1} .
\end{equation}
Here, we assumed the distance from Sgr\,A*.
If a past flare of Sgr\,A* was nearly at the Eddington luminosity  ($\sim$$10^{44}$~erg~s$^{-1}$), then it could have produced the observed RP.  
However, the flare should not be isotropic radiation, because the GCDX, which is in the vicinity of Sgr\,A*,  is not an RP but a CIE plasma (\citealt{2007PASJ...59S.245K}).
Therefore, GC South must be illuminated by collimated radiation.
In Section \ref{sec:origin}, we discussed a bipolar flow as the possible origin of GC South.
If this outflow also exhibited collimated X-ray radiation similar to blazar activity, then the plasma would become an RP.  
Indeed, the luminosity of blazars can reach as high as  $\sim$$10^{47}$~erg~s$^{-1}$.  
The {\it Fermi} bubbles suggest that a past outflow from Sgr\,A* occurred $10^{6}$~yr ago \citep{2010ApJ...724.1044S}.  
Although the epoch of the $Fermi$ bubbles activity is significantly older than the time scale of GC South, 
the possibility of similar activity $10^{5}$~yr ago may cannot be excluded. 

\section{Summary}
We observed the region 1\arcdeg.0--1\arcdeg.8 south of the GC with {\it Suzaku} and discovered an RP, which is not predicted by the standard shock-heated plasma evolution.
The plasma is most likely to be located in the GC region from its absorption column density.
Its size was then estimated to be 97~pc $\times$ 37~pc, and its thermal energy reaches $1.6\times10^{51}$~erg.
This plasma is unique in both its RP and its large thermal energy.

We consider that the origin of the plasma is  putative past activity in the GC region.
One possibility is small superwind due to a past starburst; a large plasma was blown out from the GC, and adiabatic expansion of the plasma formed the RP.
Another scenario is an outflow from Sgr\,A*; a plasma was formed by shock heating, and then photoionized by collimated X-rays.

\acknowledgments
The authors thank all of the {\it Suzaku} team members for  developing hardware and software, spacecraft operations, and instrument calibrations.
S.N. and H.U. are supported by the Japan Society for the Promotion of Science (JSPS) Research Fellowship for Young Scientists. 
This work is supported by JSPS Scientific Research grant numbers 20340043 and 23340047 (T.G.T.), 23000004 and 24540229 (K.K.), and 24740123 (N.M.).

%\bibliographystyle{apj}
%\bibliography{apj-jour,MyBib}

\end{document}